  \documentclass{article}

\newtheorem{defn}{Definition}

\newtheorem{teor}[defn]{Theorem}

\newtheorem{lema}[defn]{Lemma}
\newtheorem{rema}[defn]{Remark}
\newtheorem{coro}[defn]{Corollary}
\newtheorem{prop}[defn]{Proposition}

\makeatother

\font\ddpp=msbm10  at 11 truept 
\def\R{\hbox{\ddpp R}}

\def\Z{\hbox{\ddpp Z}}
     
\def\N{\hbox{\ddpp N}}

\newcommand{\Tau}{{\cal T}}

\newcommand{\cvd}{{\rule{0.5em}{0.5em}}\smallskip}

\begin{document}

\title{ Smooth globally hyperbolic splittings and temporal functions}
\author{Antonio N. Bernal and Miguel S\'anchez
\\ Depto. Geometr\'{\i}a y Topolog\'{\i}a, Universidad de Granada, \\ Facultad de Ciencias, Avda. Fuentenueva s/n, E-18071 Granada, Spain}%
\maketitle

\begin{abstract}

\noindent Geroch's theorem about the splitting of globally hyperbolic spacetimes is 
a central result in global Lorentzian Geometry. Nevertheless, this result was obtained at a topological level, and the possibility to obtain a metric (or, at least, smooth) version has been controversial since its publication in 1970. In fact, this problem has remained open until a definitive proof,  recently provided by the authors. Our purpose is to summarize the history of the problem, explain the smooth and metric splitting results (including smoothability of time functions in stably causal spacetimes), and sketch the ideas of the solution.

\end{abstract}


\smallskip


\section{Introduction}

Geroch's theorem \cite{Ge} is a cut result in Lorentzian Geometry which, essentially, establishes the equivalence for a spacetime $(M,g)$ between: (A) global hyperbolicity, i.e.,  strong causality plus the compactness of $J^+(p) \cap J^-(q)$ for all $p, q \in M$, and (B) the existence of a Cauchy hypersurface $S$, i.e. $S$ is an achronal subset which is crossed exactly once by any inextendible timelike curve\footnote{In particular, $S$ is a topological hypersurface (without boundary), and it is also crossed at some point -perhaps even along a segment- by any lightlike curve.}. Even more, the proof is carried out by finding two elements with interest in its own right: (1) an onto (global) time function $t: M \rightarrow \R$ (i.e. the onto function $t$ is continuous and increases strictly on any causal curve) such that each level $t^{-1}(t_0)$, $t_0 \in \R$, is a Cauchy hypersurface and, then, (2) a global topological splitting $M \equiv \R \times S$ such that each slice $\{t_0\}\times S$ is a Cauchy hypersurface. Recall also that the existence of a time function $t$ characterizes stably causal spacetimes (those causal spacetimes  which remain causal under $C^0$ perturbations of the metric).

The possibility to smooth these topological results or continuous elements, 
have remained as an open folk question since its publication. In fact, 
Sachs and Wu  claimed in their survey on General Relativity in 1977 \cite[p. 1155]{SW}: 
\begin{quote}
This is one of the folk theorems of the subject. It is not difficult to prove that every Cauchy surface is in fact a Lipschitzian hypersurface in $M$ \cite{Pe}. However, to our knowledge, an elegant proof that his Lipschitzian submanifold can be smoothed out [to such an smooth Cauchy hypersurface] is still missing.
\end{quote}
Recall that here only the necessity to prove the smoothness of some Cauchy hypersurface $S$ is claimed but, obviously, this would be regarded as a first step towards a fully satisfactory solution of the problem, among the following three: 
\begin{itemize} \item[(i)] To ensure the existence of a (smooth) {\em spacelike} $S$ (necessarily, such an $S$ will be crossed exactly once by any inextendible causal curve). 
\item[(ii)] To find not only a time function but also a ``temporal'' one, i.e., smooth with timelike gradient (even for any stably causal spacetime). \item[(iii)] To prove that any globally hyperbolic spacetime admits a smooth splitting $M=\R\times S$ with Cauchy hypersurfaces slices $\{t_0\}\times S$ orthogonal to $\nabla t$ (and, thus, with a metric without cross terms between $\R$ and $S$). 
\end{itemize}
Among the concrete applications of  (i), recall, for example: (a) Cauchy hypersurfaces are the natural regions to pose (smooth!) initial conditions for hyperbolic equations, as Einstein's ones, 
or (b) differentiable achronal hypersurfaces (as those with prescribed mean curvature \cite{EH}) can be regarded as  differentiable graphs on any smooth Cauchy hypersurface. The smoothness of a time function $t$ claimed in (ii), would  yield the possibility to use its gradient, which can be used to split any  stably causal spacetime, as in \cite{GK}.  
The applications of the full smooth splitting result (iii) include topics such as Morse Theory for lightlike geodesics \cite{Uh}, quantization \cite{Fu} or the possibility to use variational methods \cite[Chapter 8]{Ma}; it also opens the possibility to strengthen other topological splitting results \cite{Ne} into smooth ones.

Recently, we have given a full solution to these three questions (i)---(iii) \cite{BSa, BSb}. Our purpose in this talk is, first, to summarize the history of the problem and previous attemps (Section \ref{s2}) as well as the background results (Section \ref{s3}). In the two following  sections, our main results are stated and  the ideas of the proofs sketched\footnote{The results of Section \ref{s5} has been obtained just some weeks ago, and they 
were not known at the time of the meeting in November 2003, but they are sketched here because of their obvious interest for our problem.}. Concretely, Section \ref{s4}  is devoted to the construction of a smooth and spacelike $S$, following \cite{BSa}, and Section \ref{s5} to the full splitting of globally hyperbolic spacetimes, plus the existence of temporal functions in stably causal ones, following \cite{BSb}. The reader is referred to the original references  \cite{BSa, BSb} for detailed proofs and further discussions. 

\section{A brief history of time functions} \label{s2}

As far as we know, the history of the smoothing splitting theorem can be summarized as follows.
\begin{enumerate}
\item Geroch published his result in 1970 \cite{Ge}, stating clearly all the results at a topological level. Penrose cites directly Geroch's paper in his book (1972), and regards explicitly the result as topological  \cite[Theorems 5.25, 5.26]{Pe}. A subtle detail about  his statement of the splitting result  \cite[Theorem 5.26]{Pe} is the following. It is said there that, fixed $x\in S$, the curve $t \rightarrow \gamma(t)=(t,x) \in \R \times S$  is timelike. Recall that this curve does satisfy $t<t' \Rightarrow \gamma(t) << \gamma(t')$, but it is not necessarily smooth (its parameter is only continuous). In Penrose's definition \cite[pp. 2, 3]{Pe} timelike curves are assumed smooth, but later on \cite[p. 17]{Pe} the definition of causal curves is extended; in general, they will be just Lipschitzian, as previous curve $\gamma$.

\item Seifert's thesis (1968) develops a smoothing regularization procedure for time functions, which would yield the global splitting (Theorem \ref{t} below). His smoothing results were published in 1977, \cite{Se}, but the paper contains major gaps which could not be filled later. 

\item In Hawking, Ellis' book  (1973), the equivalence between stable causality and the existence of a (continuous) time function is achieved, by following a modification of Geroch's technique. Nevertheless, they assert \cite[Proposition 6.4.9]{HE} that stable causality holds if and only if a (smooth) function with timelike gradient exists. Unfortunately, they refer for the details of the smoothing result to Seifert's thesis. Even more, in 
\cite[Proposition 6.6.8]{HE},  Geroch's result is stated at a topological level, but they refer to the possibility  to smooth the result at the end of the proof. Nevertheless, again, the cited technique is the same for time functions.

\item In 1976, Budic and Sachs carried out a smoothing for {\em deterministic} spacetimes. One year later Sachs and Wu \cite{SW} posed the smoothing problem as a folk topic in General Relativity, in the above quoted paragraph.

\item The prestige and fast propagation of some of the previous references, made even the strongest splitting result be cited as proved in many references, including new influential references  or books (for example, \cite{Fu, Ma, Uh, Wa}). But this is not the case for most references in pure Lorentzian Geometry, as O'Neill's book \cite{O} (or, for example, \cite{BILY, EH, Ga, Ne}). Even more, in Beem, Ehrlich's book (1981) Sachs and Wu's claim is referred explicitly 
\cite[p. 31]{BE}. 

\item In 1988, Dieckmann claimed to prove the ``folk question''; nevertheless, he cited Seifert's at the crucial step \cite[Proof of Theorem 1]{Di}. More precisely, his study (see \cite{Dib}) clarified other point in Geroch's proof, concerning the existence of an appropiate finite measure on the manifold. Even though the straightforward way to construct this measure in Hawking-Ellis' book \cite[proof of Proposition 6.4.9]{HE} is correct, neither these authors nor Geroch considered the necessary abstract properties that such a measure must fulfill (in particular, 
the measure of the boundaries $\partial I^+(p)$ must be 0). Under this approach, on one hand, the admissible measures for the proof of Geroch's theorem are characterized and, on the other,  
a striking relationship between continuity of volume functions and reflectivity is obtained. 

In the 2nd edition of Beem-Ehrlich's book, in collaboration with Easley (1996), these  improvements by Dieckmann are stressed, but Geroch's result is regarded as topological, and the reference to Sachs and Wu's claim is maintained \cite[p. 65]{BEE}.

\end{enumerate}
\noindent In general, continuous functions can be approximated by smooth functions. Thus, a natural way to proceed would be to approximate the continuous time function provided in Geroch's result, by a smooth one. Nevertheless, this intuitive idea has difficulties to be formalized. Thus, our approach has been different. First, we managed to smooth a Cauchy hypersurface \cite{BSa} and, then, we constructed the full time function with the required properties \cite{BSb}.

\section{Setup and previous results} \label{s3}

Detailed proofs of the fact that the existence of a Cauchy hypersurface implies global hyperbolicity can be found, for example, in \cite{Ge, HE, O}. We will be interested in the converse, and then Geroch's results can be summarized in Theorem \ref{tt}, plus  Lemma \ref{ro} and Corollary \ref{cc}. 

\begin{lema} \label{ro}
Let $M$ be a ($C^k$-)spacetime which admits a $C^r$-Cauchy hypersurface $S$, $r\in \{0, 1, \dots k\}$. Then  $M$ is $C^r$-diffeomorphic to $\R \times S$ and all the $C^r$-Cauchy hypersurfaces are $C^r$ diffeomorphic.
\end{lema}
This lemma is proven by moving $S$ with the flow $\Phi_t$ of any complete timelike vector field. Thus, the (differentiable) hypersurfaces at constant $t \in \R$ are not necessarily Cauchy nor even spacelike, except for $t=0$.

\begin{teor}\label{tt} Assume that the spacetime $M$ is globally hyperbolic. Then there exists a continuous and onto map $t: M \rightarrow \R $ satisfying:

    (1) $S_a := t^{-1}(a)$ is a Cauchy hypersurface, for all $a\in \R$.

    (2) $t$ is strictly increasing on any causal curve.
\end{teor}
Function $t$ is constructed as 
$$
t(z)= {\rm ln} \left( {\rm vol}(J^-(z))/{\rm vol}(J^+(z))\right) 
$$ 
for a (suitable) finite measure on $M$ and, thus, global hyperbolicity  implies just its continuity. Finally, combining both previous results, 
\begin{coro} \label{cc}
Let $M$ be a globally hyperbolic spacetime. 
Then there exists a homeomorphism
\begin{equation} \label{ed1}
\Psi: M\rightarrow \R \times S_0 , \quad z \rightarrow (t(z), \rho(z)),
\end{equation}
which satisfies:

(a) Each level hypersurface $S_t = \{ z\in M: t(z)=t \}$ is a  Cauchy hypersurface.

(b) Let $\gamma_x: \R \rightarrow M$ be the curve in $M$ characterized by:
$$ \Psi (\gamma_x (t)) = (t, x) , \quad \quad \forall t \in \R .$$
Then the continuous curve $\gamma_x$ is timelike in the following sense:
$$ t < t' \Rightarrow \gamma_x(t) << \gamma_x(t').$$ 
\end{coro} 
\begin{rema}\label{r3.4} {\rm
If function $t$ in Corollary \ref{cc} were smooth with timelike gradient, then the spacetime $(M,g)$ would be isometric to $\R \times {\cal S}, \langle \cdot , \cdot \rangle = - \beta\,dt^2 + \bar g,$ where $\bar g$ is a (positive definite) Riemannian metric on each slice $t =$ constant. The splitting is then obtained projecting $M$ to a fixed level hypersurface by means of the flow of $\nabla t$. In what follows, a smooth function $\Tau$ with past-pointing timelike gradient $\nabla \Tau$ will be called {\it temporal} (and it is necessarily a time function).}
\end{rema}
For the smoothing procedure, some properties of Cauchy hypersurfaces will be needed. Concretely, by using a result on intersection theory, the following one can be proven   \cite[Section 3]{BSa}, \cite[Corollary 2]{Ga}:
\begin{prop}\label{pp}
Let $S_1, S_2$ be two Cauchy hypersurfaces of a globally hyperbolic spacetime with $S_1 << S_2 $ (i.e., $S_2 \subset I^+(S_1)$), and $S$ be a connected closed spacelike hypersurface (without boundary):

(A) If $S_1 << S$ then $S$ is achronal, and a graph on all $S_1$ for the descompositions in Corollaries \ref{ro}, \ref{cc}. 

(B) If $S_1 << S << S_2$ then $S$ is a Cauchy hypersurface.
\end{prop}

\section{Smooth spacelike Cauchy hyperfurfaces} \label{s4}
In this section, we sketch the proof of:
\begin{teor} \label{tsuave}
Any globally hyperbolic spacetime admits a smooth spacelike Cauchy hypersurface $S$.
\end{teor}
(In what follows, ``smooth'' will mean with the same order of differentiability of the spacetime). 

From Proposition \ref{pp}, given two Cauchy hypersurfaces 
$S_1 << S_2$   as in Theorem \ref{tt} (with $S_{t_i} \equiv S_i $; $ t_1<t_2$), it is enough to contruct a connected closed spacelike hypersurface $S$ with $S_1 << S << S_2$. And, in order to prove this, it suffices:

\begin{prop} \label{global}
Let $M$ be a globally hyperbolic spacetime with topological splitting $\R \times S$  as in Corollary \ref{cc}, and fix
$S_1<<S_2$. Then there exists a smooth function
$$
h: M \rightarrow [0, \infty)
$$
which satisfies:

\begin{itemize}
\item[1. ] $h(t,x)=0$ if $t\leq t_1$.
\item[2. ]  $h(t,x) > 1/2$ if $t = t_2$. 
\item[3. ] The gradient of $h$ is timelike and past-pointing on the open subset 
$V=h^{-1}((0,1/2)) \cap I^-(S_{2})$.

\end{itemize}
\end{prop}

\noindent In fact, recall that, given such a function $h$, each $s\in (0,1/2)$ is a regular value of the restriction of $h$ to $J^-(S_2)$. Thus, $S^h_s:= h^{-1}(s) \cap J^-(S_{2})$ is a closed smooth spacelike hypersurface with $S_1 << S^h_s << S_2$ and, then, a Cauchy hypersurface (in principle, Proposition \ref{pp}(B) can be applied to any connected component of $S_s^h$, but one can check that, indeed, $S_s^h$ is connected).

The construction of function $h$ is carried out in two closely related steps. The first one is a local step: to construct, around each $p \, (\in S_2)$, a function $h_p$ with the suitable properties, stated in Lemma \ref{local}. The second step is to construct (global) function $h$ from the $h_p$'s. This function will be constructed directly as a sum\footnote{In Riemannian Geometry, global objects are constructed frequently from local ones by using partitions of the unity. Nevertheless, the causal character of the gradient of functions in the partition are, in principle, uncontrolled. Then, the underlying idea to construct $h$ is to use the paracompactness of $M$ (which is implied by the existence of a Lorentzian --or semi-Riemannian-- metric) avoiding to use a partition of the unity. } $h= \sum h_i$ for suitable $h_i \equiv h_{p_i}$. This fact must be taken into account for the properties of the $h_p$'s in the first step:

\begin{lema} \label{local}
Fix $p \in S_{2}$, and a convex neighborhood of $p$, ${\cal C}_p \subset
I^+(S_{1})$ (that is, ${\cal C}_p$ is a normal starshaped neighborhood of any of its points). 

Then there exists a smooth function
$$
h_p: M \rightarrow [0, \infty)
$$
which satisfies:

(i) $h_p(p)=1$.

(ii) The support of $h_p$ (i.e., the closure of $h_p^{-1}(0, \infty)$)  is compact and included in ${\cal C}_p \cap I^+(S_1)$.

(iii) If $q \in J^-(S_{2}) $ and $h_p(q) \neq 0$ then $\nabla h_p(q)$ is timelike and past-pointing.
\end{lema}
{\em Sketch of proof.} Function $h_p$ is taken in a neighbourhood of ${\cal C}_p \cap J^-(S_2)$ as:
$$
h_p(q) = e^{d (p',p)^{-2}} \; \cdot \, e^{-d(p',q)^{-2}} , 
$$
where $d$ is the time-separation (Lorentzian distance) on ${\cal C}_p$, and $p'$ is a fixed suitably chosen point in the past of $p$. 

\cvd

\noindent Now, the second step is carried out by taking advantage  directly of the paracompactness of the manifold. Concretely, function $h = \sum_i h_i$ is obtained by choosing the $h_i$'s from the following lemma, with the $W_\alpha$'s equal to $ h_p^{-1}(1/2, \infty ),$
and  $p\in S_2$ (see \cite{BSa} for details):

\begin{lema} \label{topol}
Let $d_R$ be the distance on $M$ associated to any auxiliary complete Riemannian metric $g_R$.  Let $S_2$ be a closed subset of $M$ and ${\cal W}= \{ W_\alpha, \alpha\in {\cal I} \}$ a collection of open subsets of $M$ which cover $S_2$. Assume that each $W_\alpha$ is included in an open subset ${\cal C}_\alpha$ and the $d_R$-diameter of each ${\cal C}_\alpha$ is smaller than 1. Then there exist a subcollection ${\cal W}'=\{ W_j: j\in \N\} \subset {\cal W}$ which covers $S_2$ and is locally finite (i.e., for each $p\in \cup_j W_j$ there exists a neighborhood $V$ such that $V\cap W_j = \emptyset$ for all $j$ 
but a  finite set of indexes). Moreover, the collection $\{ {\cal C}_j: j\in \N\}$ (where each $W_j \in {\cal W}'$ is included in the corresponding ${\cal C}_j$) is locally finite too.
\end{lema}

\section{Temporal functions and the full splitting} \label{s5}
Now, our aim is to sketch the proof of the following theorem.
\begin{teor} \label{t} 
Let $(M,g)$ be a globally hyperbolic spacetime. Then, it is isometric to the smooth product manifold
$$
\R \times {\cal S}, \quad \langle \cdot , \cdot \rangle = - \beta\,d\Tau^2 + \bar g 
$$
where ${\cal S}$ is a smooth spacelike Cauchy hypersurface, $\Tau: \R\times {\cal S} \rightarrow \R$ is the natural projection, $\beta:\R \times {\cal S} \rightarrow (0,\infty)$ a smooth function, and $\bar g$ a 2-covariant symmetric tensor field on $\R \times {\cal S}$, satisfying:

\begin{enumerate}
\item $\nabla \Tau$ is timelike and past-pointing on all $M$, that is, function $\Tau$ is temporal. 

\item Each hypersurface ${\cal S}_\Tau$ at constant $\Tau$ is a Cauchy hypersurface, and 
the restriction $\bar g_{\Tau}$ of $\bar g$ to such a ${\cal S}_\Tau$ is a Riemannian metric (i.e. ${\cal S}_\Tau$ is spacelike).

\item The radical of $\bar g$ at each $w\in \R \times {\cal S}$ is Span$\nabla \Tau$ (=Span $\partial_\Tau$) at $w$.

\end{enumerate}
\end{teor}
Essentially, it is enough for the proof to obtain a temporal function $\Tau : M \rightarrow \R$ 
such that each level hypersurface is  Cauchy, see Remark \ref{r3.4}. The existence of such a $\Tau$ is carried out in three steps.

\smallskip 

\noindent {\em Step 1: time step functions would solve the problem.} 
Let $t \equiv t(z)$ be a continuous time function as in Geroch's Theorem \ref{tt}. Fixed $t_-<t \in \R$, we have proven in Section \ref{s4}  the existence of a smooth Cauchy hypersurface ${\cal S}$ contained in    $t^{-1}(t_-, t)$; this hypersurface is obtained as the regular value of certain function  $h\equiv h_t$ with timelike gradient on $t^{-1}(t_-, t]$. As $t_-$ approaches $t$, ${\cal S}$ can be seen as a smoothing of $S_{t}$; nevertheless ${\cal S}$ always lies in $I^-(S_t)$. Now, we claim that  the required splitting of the spacetime would be obtained if we could strengthen the requirements on this function $h_t$, ensuring the existence of  a {\em time step function} $\tau_t$ around each $S_t$. Essentially such a $\tau_t$ is a function with timelike gradient on a neighborhood of $S_t$ (and 0 outside) with level Cauchy hypersurfaces which cover a rectangular neighbourhood of $S_t$:

\begin{lema} \label{ltsf}
All the conclusions of Theorem \ref{t} will hold if the globally hyperbolic spacetime $M$ admits, around each Cauchy hypersurface $S_t$, $t\in \R$, a (time step) function $\tau_t: M\rightarrow \R$ which satisfies:
\begin{enumerate}

\item $\nabla \tau_t$ is timelike and past-pointing where it does not vanish, that is, in the interior of its support
$V_t :={\rm Int(Supp}(\nabla \tau_t)$).

\item $-1 \leq \tau_t \leq 1$.

\item $\tau_t(J^+(S_{t+2})) \equiv 1$, $\tau_t(J^-(S_{t-2})) \equiv -1$. 

\item $S_{t'} \subset V_t $, for all $t' \in (t-1,t+1)$; that is, the gradient of $\tau_t$ does not vanish in the rectangular neighborhood of $S$, $t^{-1}(t-1,t+1) \equiv (t-1,t+1)\times S$.

\end{enumerate}

\end{lema}
{\em Sketch of proof.} Consider such a function $\tau_k$ for $k\in \Z$, and define the (locally finite) sum $ \Tau = \tau_{0} + \sum_{k=1}^{\infty} (\tau_{-k} + \tau_k )$. One can check that $\Tau$ fulfills the required properties in Remark \ref{r3.4}, in fact: (a) $\Tau$ is temporal because subsets $V_{t=k}, k\in \Z$ cover all $M$ (and the timelike cones are convex), and (b) the level hypersurfaces of $\Tau$ are Cauchy  because, for each inextendible timelike curve $\gamma: \R \rightarrow M$ parameterized with $\Tau$, 
lim$_{s\rightarrow \pm \infty}(\Tau(\gamma(s))) = \pm \infty$. 

\cvd

\noindent {\em Step 2: constructing a weakening of a time step function.} Lemma \ref{ltsf} reduces  the problem to the construction of a time step function $\tau_t$ for each $t$. We will start by constructing a function $\hat \tau_t$ which satisfies all the conditions in that lemma but the last one, which is replaced by:
\begin{itemize}
\item[$\hat 4$.] $S_t\subset V_t$.
\end{itemize}
The idea for the construction of such a $\hat \tau_t$ is the following. Consider function $h$ in Lemma \ref{global} for $t_1 = t-1, t_2 = t$. From its explicit construction, it is straightforward to check that $h$ can be also assumed to satisfy: $\nabla h$ is timelike and past-pointing on a neighborhood $U' \subset I^-(S_{t+1})$ of $S_t$. Thus, putting $U= U' \cup I^-(S_t)$ ($U$ satisfies $J^-(S_{t-1}) \subset U \subset I^-(S_{t+1}) $),
we find a function $h^+: M\rightarrow \R$ which satisfies:

(i$^+$) $h^+\geq 0 $ on $U$, with $h^+ \equiv 0 $ on $I^-(S_{t-1})$. 

(ii$^+$) If $p\in U$ with $h^+(p) >0$ then $\nabla h^+(p)$ is timelike and past-pointing.

(iii$^+$) $h^+ > 1/2 $ (and, thus, its gradient is timelike past-pointing) on $J^+(S_t)\cap U$.

\smallskip

\noindent Even more, a similar reasoning  yields a 
 function $h^-: M\rightarrow \R$ for this same $U$ which satisfies:

(i$^-$) $h^- \leq 0$, with $h^- \equiv 0$ on $M\backslash U$.

(ii$^-$) If $\nabla h^-(p)\neq 0$ at $p \, (\in U)$ then $\nabla h^-(p)$ is timelike past-pointing.

(iii$^-$) $h^- \equiv -1$ on $J^-(S_t)$.

\smallskip \noindent Now, as $h^+ - h^- >0$ on all $U$, we can define: 
$$
\hat \tau_t = 2 \; \frac{h^+}{h^+ - h^-} -1
$$
on $U$, and constantly equal to 1 on $M\backslash U$. A simple computation shows  that $\nabla \hat \tau_t$ does not vanish wherever either $h^- \nabla h^+$ or $h^+ \nabla h^-$ does not vanish (in particular, on $S_t$) and, then, it fulfills all the required conditions. 

\smallskip

\noindent {\em Step 3: construction  of a true time step function.} Now, our aim is to obtain a function $\tau (\equiv \tau_t)$ which satisfies not only the requirements of previous $\hat \tau \equiv \hat \tau_t$ but also the stronger condition 4 in Lemma \ref{ltsf}. Fix any compact subset $K \subset t^{-1}([t-1,t+1])$. From the construction of $\hat \tau$, it is easy to check that $\hat \tau$ can be chosen with $\nabla \hat \tau$ non-vanishing on $K$. Now, choose a sequence $\{G_j: j \in \N \}$ of open subsets such that:
$$
\overline{G_j}\,\,\,\mbox{is compact},\,\,\,\overline{G_j}\subset G_{j+1}
 \,\,\,\mbox{}\,\,\,M=\cup_{j=1}^\infty G_j , 
$$ 
and let $\hat \tau [j]$ be the corresponding sequence of functions type $\hat \tau$ with gradients non-vanishing on:
$$
K_j=\overline{G_j}\cap t^{-1}([t-1,t+1]) .$$ 
Essentially, the required time step function is:
$$ \tau = \sum_{j=1}^\infty \frac{1}{2^j A_j} \hat \tau [j],
$$
where each $A_j$ is chosen to make $\tau $  smooth (fixed a locally finite atlas on $M$, each $A_j$ bounds on $\overline{G_j}$ each function $\hat \tau [j]$ and its partial derivatives  up to order $j$ in the charts of the atlas which intersect $G_j$). Then, the gradient of $\tau$ is timelike  wherever one of the gradients $\nabla \hat \tau [j]$ does not vanish (in particular, on 
$t^{-1}([t-1,t+1])$). Moreover, $\tau$ is equal to constants (which can be rescaled to $\pm 1$) on $t^{-1}((-\infty, t-2]), t^{-1}([t+2,\infty))$, as required.
\cvd 


\noindent Finally, it is worth pointing out that similar arguments work to find a smooth time function on any spacetime (even non-globally hyperbolic) which admits a continuous time function $t$.

\begin{teor} \label{t0b}
Any  spacetime $M$ which admits a (continuous) time function (i.e., is stably causal) also admits a temporal function $\Tau$. 
\end{teor}
{\em Sketch of proof.}  Notice that each level continuous hypersurface $S_t$ is a Cauchy hypersurface in its Cauchy development $D(S_t)$. Moreover, any time step function $\tau_t$ around $S_t$ in $D(S_t)$ can be extended to all $M$ (making $\tau_t $ equal to 1 on $ I^+(S_t) \cap (M\backslash D(S))$, and to $-1$ on $ I^+(S_t) \cap (M\backslash D(S))$). Then, sum suitable time step functions as in Step 3 above.

\cvd





\section*{Acknowledgments}

The second-named author has been partially supported by a  MCyT-FEDER Grant BFM2001-2871-C04-01.

  \end{document}